\documentclass[fleqn,10pt]{wlscirep}
\usepackage[utf8]{inputenc}
\usepackage[T1]{fontenc}

\makeatletter
\newcommand*{\rom}[1]{\expandafter\@slowromancap\romannumeral #1@}
\makeatother
\title{Speckle Tweezers for Simultaneous Controlling of  Low and High Refractive Index Micro-particles}

\author[1]{Ramin Jamali}
\author[2]{Farzaneh Nazari}
\author[3]{Azadeh Ghaffari}
\author[4]{Sabareesh K. P. Velu}
\author[1,5,*]{Ali-Reza Moradi}

\affil[1]{Department of Physics, Institute for Advanced Studies in Basic Sciences (IASBS), Zanjan 45137-66731, Iran}
\affil[2]{Department of Physics, Yazd University, Yazd 89195-741,  Iran}
\affil[3]{Department of Food and Drug Control, School of Pharmacy, Zanjan University of Medical Sciences, Zanjan 45139-56111, Iran}
\affil[4]{Department of Physics, Rathinam College of Arts and Science,  Coimbatore 641021, Tamilnadu, India\\
Email: dean.rd@rathinam.in}
\affil[5]{School of Nano Science, Institute for Research in Fundamental Sciences (IPM), Tehran 19395-5531, Iran}

\affil[*]{moradika@iasbs.ac.ir}

\begin{abstract}
Manipulation of micro and nanoscale particles suspended in a fluidic medium is one among the defining goals of modern nanotechnology. Speckle tweezers (ST) by incorporating randomly distributed light fields have been used to apply detectable limits on the
 Brownian motion of micro-particles with refractive indices higher than their medium. Indeed, compared to periodic potentials, ST
 represents a wider possibility to be operated for such tasks. Here, we extend the usefulness of ST into low index micro-particles. Repelling of such particles by high intensity regions into lower intensity regions makes them to be locally confined, and the confinement
 can be tuned by changing the average grain intensity and size of the speckle patterns. Experiments on polystyrenes and liposomes
 validate the procedure. Moreover, we show that ST can be used to nano-particle (NP)-loaded liposomes. Interestingly, the different interactions of NP-loaded and empty liposomes with ST enables collective manipulation of their mixture using the same speckle
 pattern, which may be explained by inclusion of the photophoretic forces on NPs in NP-loaded liposomes. Our results on the different behavior between empty and non-empty vesicles may open a new window on controlling collective transportation of drug micro-containers along with its wide applications in soft matter.
\end{abstract}
\begin{document}

\flushbottom
\maketitle
%
%

\section{Introduction}
Using light to control the motion of micro- and nano-structured objects is a  challenge and involves several scientific and technological fields such as optical tweezing \cite{grier2003revolution}, Van der Waals and Casimir interactions \cite{hertlein2008direct,munday2010repulsive}, integrated optics \cite{tien1971light}, biophysics \cite{campbell2012introduction}, etc. However, in  more complex light-activated devices the unavoidable   disorder   induces  some  effects including multiple-scattering, diffusion, and  localization of light \cite{wiersma2013disordered,rotter2017light,jacucci2020light}. The disorder can be externally controlled, which leads to applications such as tunable  lasers, transmission of light through random media, and novel disorder driven devices, such as ultrasensitive
spectrometers \cite{wiersma2001temperature,redding2013compact,rotter2017light}.  Yet, there is a specific field -optomechanics-  in which  the beneficial features of randomness have not  been sufficiently  investigated. Optomechanical forces come from the interaction of the electromagnetic  wave with the boundaries of dielectric objects. On the other hand, random systems include  a large number of boundaries for which calculation of the  forces is not trivial, but  it is crucial to understand the mechanisms of optically activated devices \cite{eichenfield2007actuation}.

Manipulation of micro and nanoscale particles suspended in a fluidic medium, in general, is one of the defining goals of modern nano-technology \cite{grigorenko2008nanometric}. The conventional strategy is to trap particles utilizing optical trapping forces \cite{grier2003revolution}, photo-acoustic effects  \cite{shi2009acoustic}, and magnetic  and electric forces  \cite{fan2011electric}. Progressions into a specific approach has resulted in numerous breakthroughs in pharmaceutics and microfluidics \cite{nilsson2009review,yanik2011technologies,ghosh2018mobile}. A primary advantage of specific methodology is the opportunity for it to be applied in the manipulation of  drug containers in targeted drug delivery \cite{spyratou2012biophotonic}.  Among various approaches, optical tweezers have been particularly prosperous due to their inherent versatility.   Optical trapping of micron-sized particles was introduced by A. Ashkin \cite{ashkin1970acceleration}. Since then, it has been successfully implemented in two size ranges: the subnanometer scale for cooling of atoms, ions and molecules \cite{verkerk1992dynamics,haroche2013nobel}, and the micrometer scale for non-invasive  manipulation of microscopic objects such as cells and bacteria \cite{curtis2002dynamic}.  Further developments through externally controlling  the position and stiffness of individual trap sites, offer manipulation tasks in parallel and lead to many important breakthroughs in bioscience, materials science, microfluidics, and soft condensed matter physics \cite{gao2017optical}. Conventional optical traps require careful engineering and aligning of  setups and sample preparation. Therefore, such stringent conditions  are not matchable with the simplicity, low-cost and high-throughput requirements that are necessary for  biomedical applications.

The light gradient force exerted on a dielectric particle must  overcome the scattering force for steady trapping, especially along the axial  direction. The gradient force acting on the particle is proportional to $\pm \nabla E^2$, where $E$ is  the electric field of the beam, and the difference in the refractive indices between the surrounding medium $n_m$ and the particle $n_p$ indicates the sign.  The realization of stable trapping of high-refractive-index particles is easy since the intensity profile of tightly focused  laser beams with a spatially homogeneous state of polarization is in the form of a Gaussian distribution. However, for low-refractive-index particles, i.e., when the refractive index of the particle is lower than that of the surrounding medium ($n_p < n_m$), the gradient force will point from the focus center to the region of low-intensity and will repel the particle away from the highest intensity point. Therefore, low-index particles require a  different strategy to be manipulated by  optical tweezers; to trap them, a beam with a focusing feature of hollow intensity distribution, the so-called hollow beam, is used. Various types of hollow beams, such as high order Bessel beams, Laguerre-Gaussian beams, azimuthally polarized beams, and the output beam of small hollow fibers, etc. have been tested and applied for this purpose in recent years \cite{miyazaki2009motion,peng2009trapping,prentice2004manipulation}.  Trapping by hollow beams, furthermore,  results in a reduction of possible photo-damage, which is crucial for biological specimens \cite{dasgupta2010optical}. 

On the other hand, it is known that low-refractive-index particles have an important role in physics, medicine, and technology. For example, trapping a single gas bubble in water in an acoustic resonant cavity has allowed  several studies to understand the physics of sonoluminescence \cite{putterman2001there}  and to establish further applications in biology and medicine. More importantly, the bubbles are used to enhance the contrast of images in ultrasound imaging \cite{de2002basic}. Besides bubbles, low-refractive-index particles are found in water-in-oil emulsions for petroleum, and food processing applications \cite{ye2004trapping}.  Another important class of  low-index particles are liposomes. A liposome is a spherical vesicle with a molecularly thin and self-assembled membrane, and may act as an appropriate container for nutrients and pharmaceutical medicines in drug delivery  \cite{discher2002polymer} and can be applied to enhance the transfection of genes into living cells \cite{mohanty2007trapping,tachibana1999induction}. Liposomes are made of phospholipids and are developed by disrupting biological membranes,  their most important kind   being the multilamellar vesicles  \cite{zarif2005drug}.  

Usually, it is difficult to optically trap both kinds of particles (low and high-refractive-indices) as it  requires complex scanning methods and/or rigorous design of holographic tweezers \cite{hansen2005expanding,rui2015manipulation,gahagan1999simultaneous,liu2020simultaneous}. Therefore, conventional methods are not suitable for more complex micro-manipulation tasks such as transportation of a large number of such mixtures of particles. Also, limitations such as the short working distance of  high numerical aperture microscope objectives  undermine the efficiency of the methods.  
In this paper, we propose the use of speckle  tweezers (ST). ST possesses both the simplicity and  applicability in optical manipulation of real-life situations. Speckle light patterns are high-contrast, fine-scale granular patterns that are the result of the interference of a large number of dephased but coherent monochromatic light waves propagating along different directions  \cite{bouchaud1990anomalous}. Speckle patterns can be generated through different processes, such as scattering of  laser light from a rough surface or mode-mixing in a multimode fiber \cite{goodman2007speckle}.  It has been    demonstrated that  ST is a versatile tool to efficiently perform collective optical manipulation tasks, such as trapping, guiding, and sorting  of high-refractive-index particles  \cite{volpe2014brownian,shvedov2010selective}.  Although, the intrinsic randomness of speckle patterns is considered to be a disturbing noise in experiments, nevertheless, the patterns find    several useful applications.  One of the first experiments on the use of ST in this range demonstrated the emergence of anomalous diffusion in colloids and showed the control of the motion  of Brownian particles \cite{chubynsky2014diffusing}. Recently, the optical speckle field has been used for producing a thermal speckle field through interaction  with plasmonic substrates and converting the high-intensity  speckle grains into the corresponding thermal speckle grains  \cite{kotnala2020opto}.  In another recent study,  it has been shown that the density of the structural defects in a 2D binary colloidal crystal can be engineered using a speckle field \cite{nunes2020ordering}. A memory equation based on a theoretical model was provided to describe the motion of colloidal particles within the ST field \cite{pincce2016disorder,recanello2020extended}. 
The speckle field  contains a random, but somehow tunable, distribution of isolated high-intensity regions surrounded by irregular and interconnected low-intensity  regions, and the dark spots are, indeed, much more common than bright ones \cite{freund19981001,berry2000phase}. This nature of the speckle fields makes  the use of ST as an elegant  possibility for collective manipulation of  mixtures of low and hight index micro-particles. 
Regions with higher and lower intensities in the speckle field can be used to confine the high-refractive-index and low-refractive-index particles, respectively. 
Moreover, by examination of speckle fields effect on not-empty but nano-particle (NP) loaded liposomes, we experimentally corroborate that the ST may be used to perform control on the collective movement of drug micro-containers.

\section{Materials and methods}
\begin{figure*}[t!]
	\begin{center}
		\includegraphics[width=.89\textwidth, angle=0]{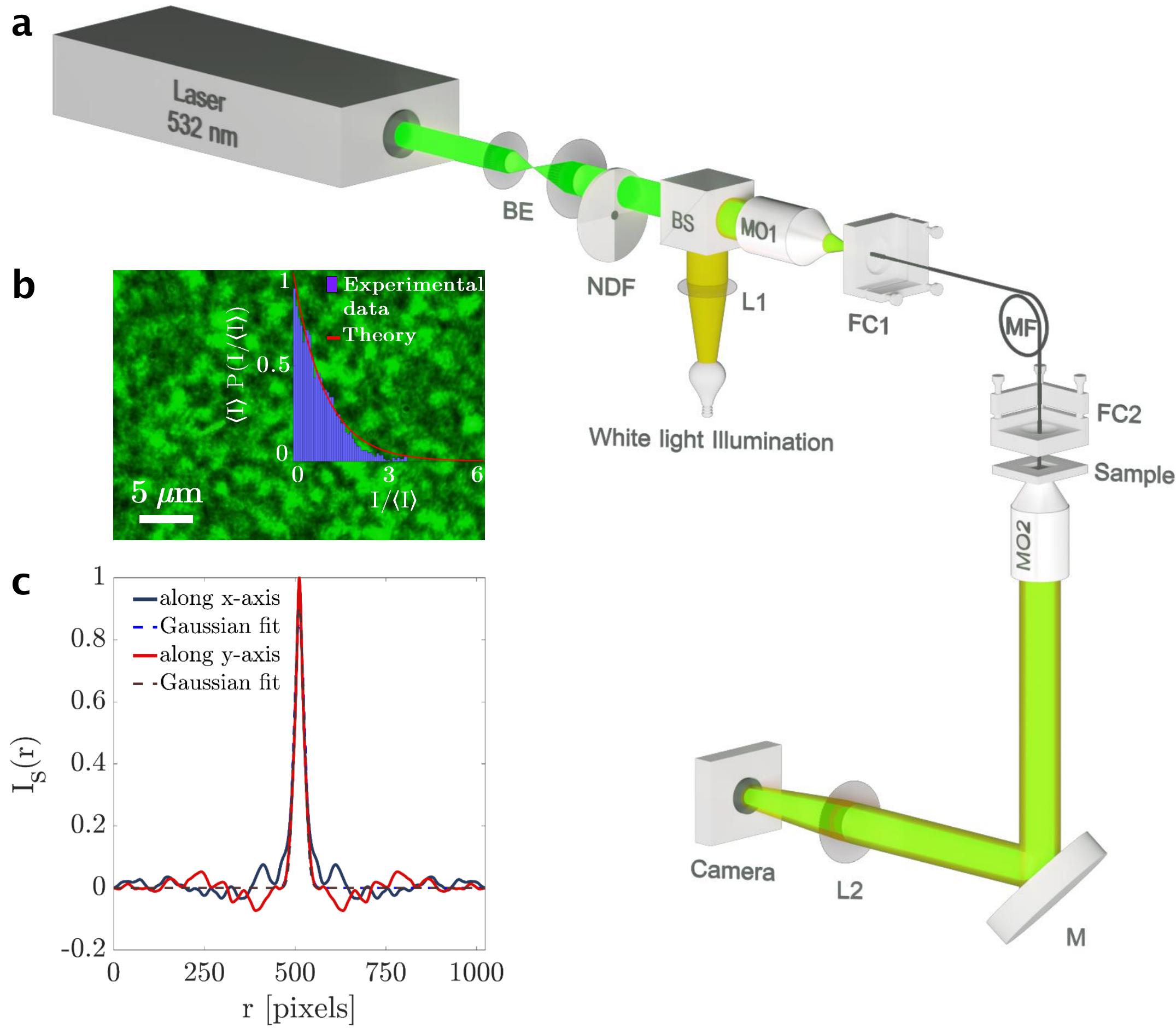}
		\caption{(a) Schematic of the experimental setup; BE: beam expander, MO: microscope objective, MF: multi-mode fiber, NDF: neutral density filter, BS: beam splitter, FC: fiber coupler, L: lens, M: mirror. (b) A bright field microscopy image of the speckle pattern at the sample plane. Inset: Intensity distribution of speckle pattern. Blue bars demonstrate the experimental data and red solid line shows the  theoretical probability density  function, which follows  the negative exponential distribution. (c) Normalized speckle pattern intensity spatial auto-correlation function $I_s({r})$ as a function of  position $r$. The blue and red lines are speckle intensity along $x$ and $y$ axis, respectively, and the dashed blue and red lines are their corresponding  Gaussian functions. The average speckle grain size is determined from the standard deviation of the Gaussian function. In our case, the average speckle grain size is 1.69 $\pm$ 0.20 $\mu$m.}
	\end{center}
	\label{Figure01}
\end{figure*}
\subsection{Sample preparation}
Liposomal of clindamycin phosphate (CP) is prepared by thin-film hydration method and  characterized in terms of size, size distribution, zeta potential, stability, and encapsulation efficiency.  CP is purchased from Suzhou Pharmaceutical. Monobasic potassium phosphate (98- 100.5\%), phosphoric acid (85-88\%), sodium hydroxide ($>$98\%), Tween 80 (for synthesis), triton X-100 (99\%) and high performance liquid chromatography HPLC-grade acetonitrile are purchase from Merck. Sephadex G-50 is purchased from MilliporeSigma. Egg phosphatidyl cholin (E80) is purchased from Lipoid. Cholestrol (99\%) is purchased from Sigma-Aldrich. The lipid mixture  is dissolved in chloroform/methanol (2:1) and get dried under a reduced pressure in a rotary evaporator at 60$^{\circ}$C  to form a thin lipid film. The film is then hydrated with CP solution (63 mg/ml) in phosphate buffer solution pH = 7.4 at 60$^{\circ}$C for 1 hour.  

The liposome size can vary from very small (0.025 $\mu$m) to large (2.5 $\mu$m) vesicles. Moreover, liposomes may have one or more  bilayer membranes. The vesicle size is an acute parameter to  determine the circulation half-life of the liposomes. We  separate out the liposomes of  1 $\mu$m range by passing them through syringe filters (Sterlitech).  These liposomes are in  multilamellar vesicles class, i.e., their  structures consist of  concentric phospholipid spheres separated by layers  of water.  The refractive index of the formed liposomes is 1.300, which is measured by Abbe refractometer  (Atago, NAR-1T).

Gold NPs of 100$\pm$5  nm diameter and optical density of 1, and polystyrene particles of 1 $\mu$m diameter are  purchased from Sigma-Aldrich. Citrate buffer, which is a proprietary surfactant is used as a stabilizer for  NPs.  Gold NPs  are loaded  passively to the liposomes i.e. encapsulated during the liposome formation. To this end, during the liposome preparation and in the hydration stage, 100 microliter of gold NPs are added to distilled water.  Before the experiments,  polystyrene particles,  liposomes,  and the NP-loaded liposomes are shaken for  a few minutes by an ultrasonic bath (Xuba3, Grant), and separate dilute solutions of them, which are  suitable for trapping experiments, are prepared. The concentration could be further adjusted by the use of a  push-pull syringe pump, connected into the surrounding medium and the high concentrate  liposome reservoirs. For example,  for increasing the concentration, we infuse the liposomes directly into the chamber and withdraw the surrounding medium at the constant rate. 

We fabricate special chambers with a couple of glass coverslips (22 $\times$ 22 mm,  thickness of 170 $\pm$ 5 $\mu$m) and parafilm spacers, in which parts of one of the building coverslips are coated by a semi-transparent layer, for reducing the transmittance of the laser light up to 15\%. The coating is performed using Sliver (Ag) sputter targets  in a sputter coater system (DST3-A, Nanostructured Coatings Co.) on the pre-covered areas of the coverslip. By tuning the coating duration and the applied voltage the transmittance of the coated area can be tuned.

\subsection{Experimental Procedure}
Figure 1\textbf{a} shows the schematic of the home-built STs experimental setup to perform the experiments.  A laser beam (DPSS, 85-GSS-309, Melles Griot, 532 nm) is expanded by the beam expander (BE) and by the use of a microscope objective (MO1, $10\times$, NA = 0.25) is focused onto the input entrance of a multi-mode fiber (MF).  A  step-variable neutral density filter (NDF) is used to reduce the beam intensity in steps. A beam splitter (BS) is also used to direct a white light illumination along with laser light into the sample for conventional microscopic imaging.  
The multi-mode fiber  has a core diameter of $365 \pm 14~\mu $m, numerical aperture of NA = 0.22, and core and cladding refractive indices of 1.4589 and 1.4422, respectively.
Therefore, our MF can mix up to approximately 474 modes of the propagating laser beam, and the output light from the fiber is the   interference of several modes with random phases and results in generating the speckle pattern at the sample plane. The microscope objective MO2 ($40\times$, NA = 0.65), mirror (M), and the lens (L2)  form the image of the sample on the camera (DCC1545M, Thorlabs, 5.2 $\mu$m pixel pitch).
The sample chamber consists of a microfluidic  channel with an inlet and outlet to inject, withdraw, or flow aqueous  samples by the use of a syringe pump (NE-300, Just Infusion\texttrademark). 

Figure 1\textbf{b} is a typical speckle pattern   as observed on the camera. The average grain size of the speckle field in this pattern is  1.69 $\mu$m and the average speckle intensity is $\langle I\rangle = 200 $mW/$\mu \rm{m}^2$. In its inset the theoretical (red color solid line) and experimental  (blue color bars) probability density function of the speckle pattern intensity is shown. 
The experimental data is obtained by averaging the intensity distribution of 1000 speckle patterns. The probability density function of the speckle pattern intensity follows the negative exponential distribution, i.e., $\frac{1}{\langle I\rangle} \exp{(-\frac{I}{\langle I\rangle})}$ \cite{goodman2007speckle,bender2018customizing}. Figure 1\textbf{b} shows a good agreement between theory and  the distributions of the speckle pattern intensities used in the experiments, which confirms  that the speckle patterns used in the experiments are well-formed.

Figure 1\textbf{c} shows the auto-correlation function of the speckle pattern intensity. The average speckle grain size is determined by evaluating the normalized spatial auto-correlation function ($I_s(r)$) on the 2D imaged speckle patterns. The resulted normalized auto-correlation function is  fitted by the Gaussian function, whose standard deviation is approximately one-third of the average speckle grain size.
\begin{figure*}[t!]
	\begin{center}
		\includegraphics[width=.8\textwidth]{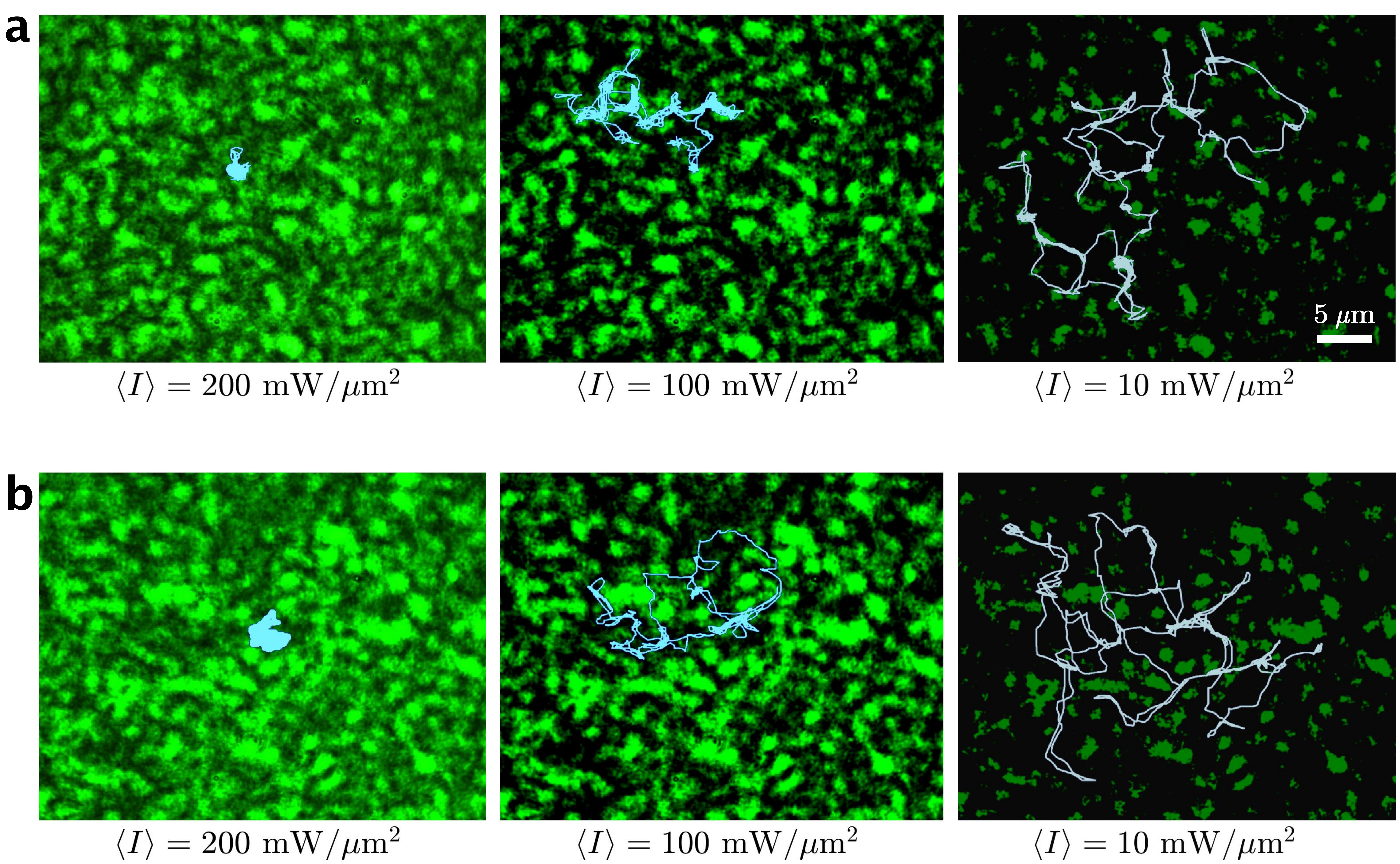}
		\caption{The experimental trajectories (turquoise solid lines) of high and low-refractive-index particles in  regressive confinement through decreasing the average speckle intensity ($\langle I\rangle = 200 $ mW/$\mu \rm{m}^2$, $\langle I\rangle = 100 $ mW/$\mu \rm{m}^2$, $\langle I\rangle = 10 $ mW/$\mu \rm{m}^2$). The surrounding medium is distilled water of refractive index $n_m = 1.33$, viscosity of $\eta_w$ = 0.001 $\rm{Ns/m^2}$  at temperature $T = 300$ K.  The background represents the speckle field generated by propagating  of a solid state laser light through a multimode fiber.  
			(a) Trajectories of a polystyrene particle of diameter $D_p = 1$ $ \mu$m and refractive index of $n_p = 1.59$.   
			(b) Trajectories of a  liposome vesicle of diameter $D_l=1$ $ \mu$m,  refractive index of $n_l = 1.3$, and viscosity of $\eta_l = 340$ $\times 10^{-3}$ $ \rm{Ns/m^{2}}$.}
		\label{Figure02}
	\end{center}
\end{figure*}
\section{Results and Discussion}
To demonstrate trapping and the manipulation of low-refractive-index particles using a speckle field we start by considering the motion behavior of two samples containing polystyrene microspheres (diameter of 1 $\mu$m, refractive index of $n_p$ = 1.59) and liposome  vesicles (diameter of 1 $\mu$m, refractive index of $n_l$ = 1.3) under the exposure of randomly distributed  laser field.  We brought the output end of a multimode fiber into the specimen chamber to impose the speckle field on the dispersed particles. The average intensity of the speckle grains is controlled by tuning the intensity of the laser light, and the grain size can be varied by careful adjustment of the fiber end to sample distance. The microfluidic chamber is connected to an infusion pump to infuse the micro-particles and allow them to flow (infusion rate=1.25 ml/hr). First, in  separate experiments, we imposed the speckle field on multiple low- and high-refractive-index particles in a quiescent medium ($n_m=1.33$). Figures 2\textbf{a} and 2\textbf{b} show the path trajectories  of the typical low- and high-refractive-index   particles, respectively, for the different speckle field strengths.  As the trajectories show, polystyrene beads  reposition themselves in the bright regions of the speckle pattern and under strong optical forces ($\langle I\rangle = 200$ mW/$\mu \rm{m}^2$) they remain trapped in one of the speckle bright grains for several minutes. However, according to their Brownian motion, the particles can also move inside the speckle grain. By reducing the average laser intensity to $\langle I\rangle = 100$ mW/$\mu \rm{m}^2$ they are still trapped but can move between the neighboring grains. By further reducing the average intensity ($\langle I\rangle = 10$ mW/$\mu \rm{m}^2$) the optical forces become relatively low and the particles may show almost free diffusion but still feel the speckle field. Similar behavior is observed in liposome vesicles. However, predictably,  according to their low-refractive-index with respect to their surrounding medium, as the trajectories of Fig. 2\textbf{b} show, the liposomes are restricted to the dark regions with high laser intensity  and  move between few neighboring dark regions with lower intensity. Eventually when the average intensity reduces to $\langle I\rangle = 10$ mW/$\mu \rm{m}^2$ the liposomes can move between several dark regions. The optical force vector acting on  a moving particle moves in the speckle field and  varies with a characteristic time scale, which is   inversely proportional to the average speckle intensity  in the first approximation \cite{volpe2014brownian}.  The motion of a Brownian particle in a static speckle field is the result of random thermal forces and deterministic optical forces. However, the optical  forces are the main  forces in these experiments. They attract high-index particles  towards the intensity maxima of the  optical field but  repel the low-index ones to the outer sides of the speckle grains resulting in trapping them in the dark regions.
The optical forces act on both high- and low-index particles stumbling in the speckle field  in a characteristic time, the so-called waiting time, which is defined as the time that the particle moves along its correlation length with its average drift speed \cite{volpe2014brownian}. The average particle drift speed is directly proportional to the average force and inversely  to its friction coefficient. It has shown that even a random optical field, regardless of its inherent randomness, may be used to control the motion of the high-index particle \cite{volpe2014brownian}.  For a low-index particle, this is performed through repelling  it by the speckle grains. 

For the particle in a static speckle pattern, when the optical forces are relatively low the particle is almost freely diffusing away from a speckle grain before  experiencing the influence of the optical forces. Increasing the forces,  a subdiffusive behavior is shown and the high (low) index particle is constrained in the bright (dark) grains.  However, the particle has small probability to escape a speckle grain and can still  move between the neighboring grains. At  strong forces, the particle remains trapped in one of the  bright  or dark speckle grains, depending on the particle refractive index.

The mean square displacement (MSD) of the particle motion is a quantitative expression to describe the aforementioned behavior of the particle inside the speckle field, and is
defined as  ${\rm{MSD}}(\Delta t) = \langle \Delta r^2 (\Delta t)\rangle  = \langle [r_j (t_0 + \Delta t) - r_j(t_0)]^2 \rangle_{j,t_{0}}$, where $r_j (t)$ denotes the position of micro-particles $j$  at time $t$, and the average is derived over all micro-particles $j$ and all onset time $t_0$. It means that, we consider both the ensemble and time average.
We obtain each MSD curve  by taking an average over 1000 different  trajectories of the micro-particles we have used (polystyrene, empty liposome and NP-loaded liposome).
\begin{figure*}[t!]
	\includegraphics[width=\textwidth]{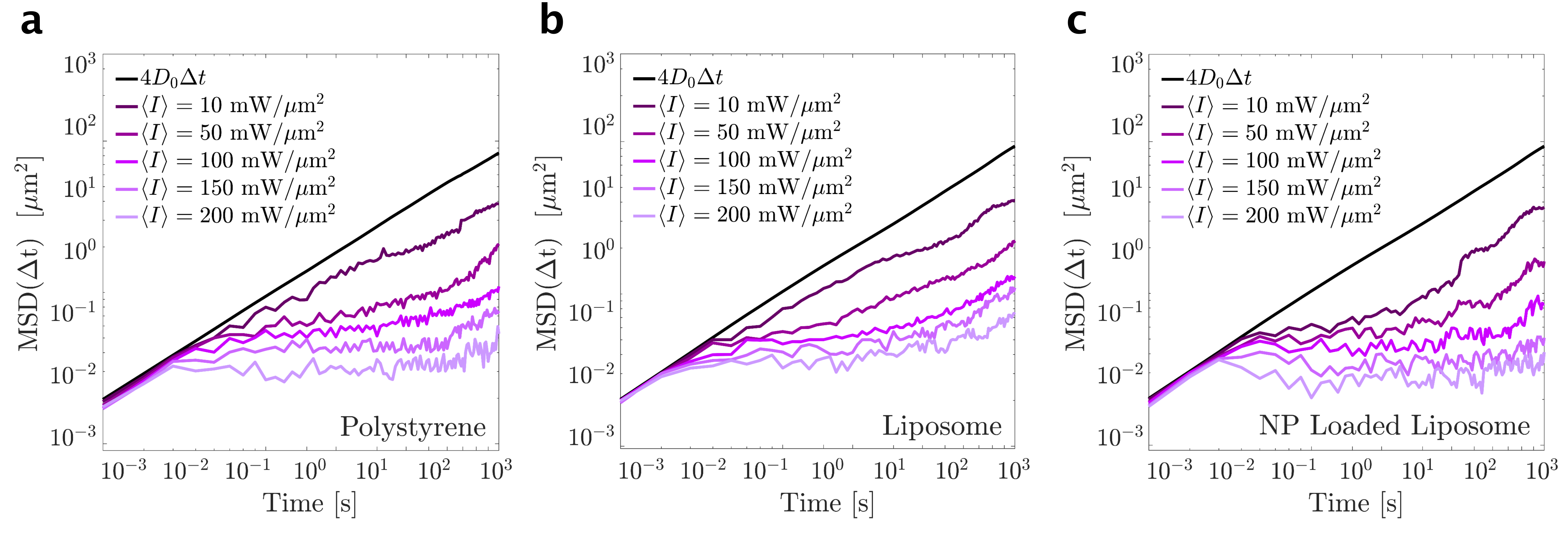}
	\caption{Mean square displacement of Brownian particles on a logarithmic scale  as a function of  time in increasing  speckle field intensity, for (a) a polystyrene, (b) a liposome, and (c) a NP-loaded liposome.  Black lines represent  Einstein's free diffusion law.   The initial position of the particles is chosen randomly and the reported  data in each of the figures are the average  of more than 1000 trajectories in a time duration of 1000 s. The average speckle intensity is increased  from 10 to 200 mW/$\mu \rm{m}^2$. In all cases, a  transition from trapping at high speckle intensities to subdiffusion at low speckle intensities  can be recognized.}
	\label{Figure03}
\end{figure*}
Figures 3\textbf{a}, 3\textbf{b}, and 3\textbf{c}  show the MSD of a polystyrene particle, a liposome, and  a  liposome containing NPs, respectively, that are moving in a static speckle field of different average speckle intensities. In the absence of optical forces or for low optical forces and for longer times, the  MSD behaves almost linearly with $\Delta t$  (MSD$\sim4D_{\rm{SE}}\Delta t$, where $D_{\rm{SE}}$ is the Stokes-Einstein diffusion coefficient). As the forces increase and $t$ decreases, there is a transition into a subdiffusive regime characterized by MSD $\propto\Delta t^{\beta} $ with $\beta < 1$. For a very large $\Delta t$, the motion returns into the diffusive regime, i.e., $\beta = 1$,  with a smaller effective diffusion coefficient $D_{\rm{eff}}$. Under higher external speckle forces, after the waiting time, the particle takes a rest time until bounded in another speckle grain. For even higher forces the particles cannot transit from one speckle grain to another. Surprisingly, this behavior and diffusion-to-subdiffusion transition are very similar for all the examined particles; the low-index ones are restricted to darker regions according to their trajectories.
\begin{figure*}[t!]
	\begin{center}
		\includegraphics[width=.85\textwidth]{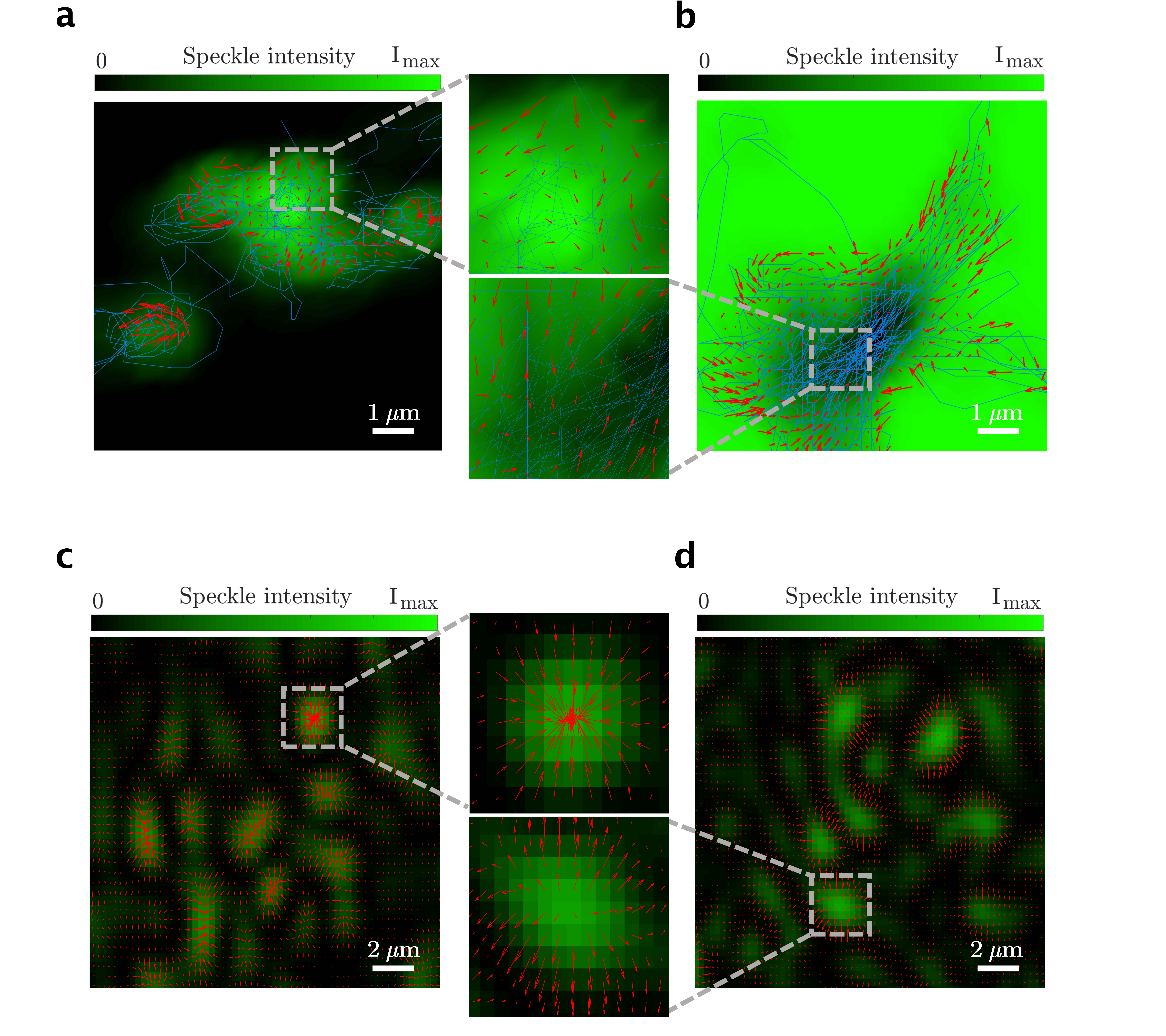}
		\caption{Microscopic optical forces in a speckle field  in the transversal plane for (a) a polystyrene (high-refractive-index,  $n_p = 1.59$) and (b) a liposome vesicle (low-refractive-index, $n_l = 1.3$), reconstructed through the analysis of particle displacement information via the maximum-likelihood-estimator analysis method. (c) and (d) The corresponding simulated force fields in speckle fields of the same average intensity as in (a) and (b).}
		\label{Figure04}
	\end{center}
\end{figure*}

To demonstrate the acting optical forces on a low-refractive-index particle in a speckle light field, we obtain the  2D microscopic force fields through force reconstruction via the maximum-likelihood-estimator analysis (FORMA) algorithm in its 2D form, which is introduced in \cite{garcia2018high}. In FORMA  the forces are retrieved by the analysis of  the particles' Brownian trajectories. The optical force fields generated by speckle patterns are  larger extended force fields and the equilibrium positions are not known \textit{a priori } due to their random appearance and include non-conservative components.  FORMA estimates accurately the conservative and non-conservative components of the force field simultaneously with important advantages over the common methods that obtain the forces by  analyzing their influence on the particles' Brownian motion. This method is more accurate, does not need  calibration fitting parameters, executes much faster,  and requires much  fewer data.  In Figs. 4\textbf{a} and 4\textbf{b} the results of the force field  reconstruction in the speckle field   are shown for the high- and low-refractive-index particles, respectively. The green color background  is the overlaid  images of the speckle field pattern, whose intensity is  proportional to the potential depth of the optical potential felt by the micro-particles. The red arrows plot the 2D force field exerted on the micro-particles and measured with FORMA. The trajectories (thin turquoise lines) are  acquired experimentally for the motion of a polystyrene particle of diameter $D_p = 1$ $ \mu$m and refractive index of $n_p = 1.59$ (Fig. 4\textbf{a}) and   liposome vesicle of diameter $D_l = 1$ $ \mu$m,  refractive index of $n_l = 1.3$, viscosity of $\eta_l = 340$ $ \times 10^{-3}$ $ \rm{Ns/m^{2}}$ (Fig. 4\textbf{b}) in distilled water of refractive index $n_m = 1.33$, viscosity of $\eta_w$ = 0.001 $\rm{Ns/m^2}$, and at temperature $T = 300$ K. As illustrated in Fig. 2, a polystyrene (a liposome) explores the high (low) intensity regions  while being trapped for a while in each bright (dark) grains before crossing over a neighbor one.
\begin{figure*}[t!]
	\includegraphics[width=\textwidth]{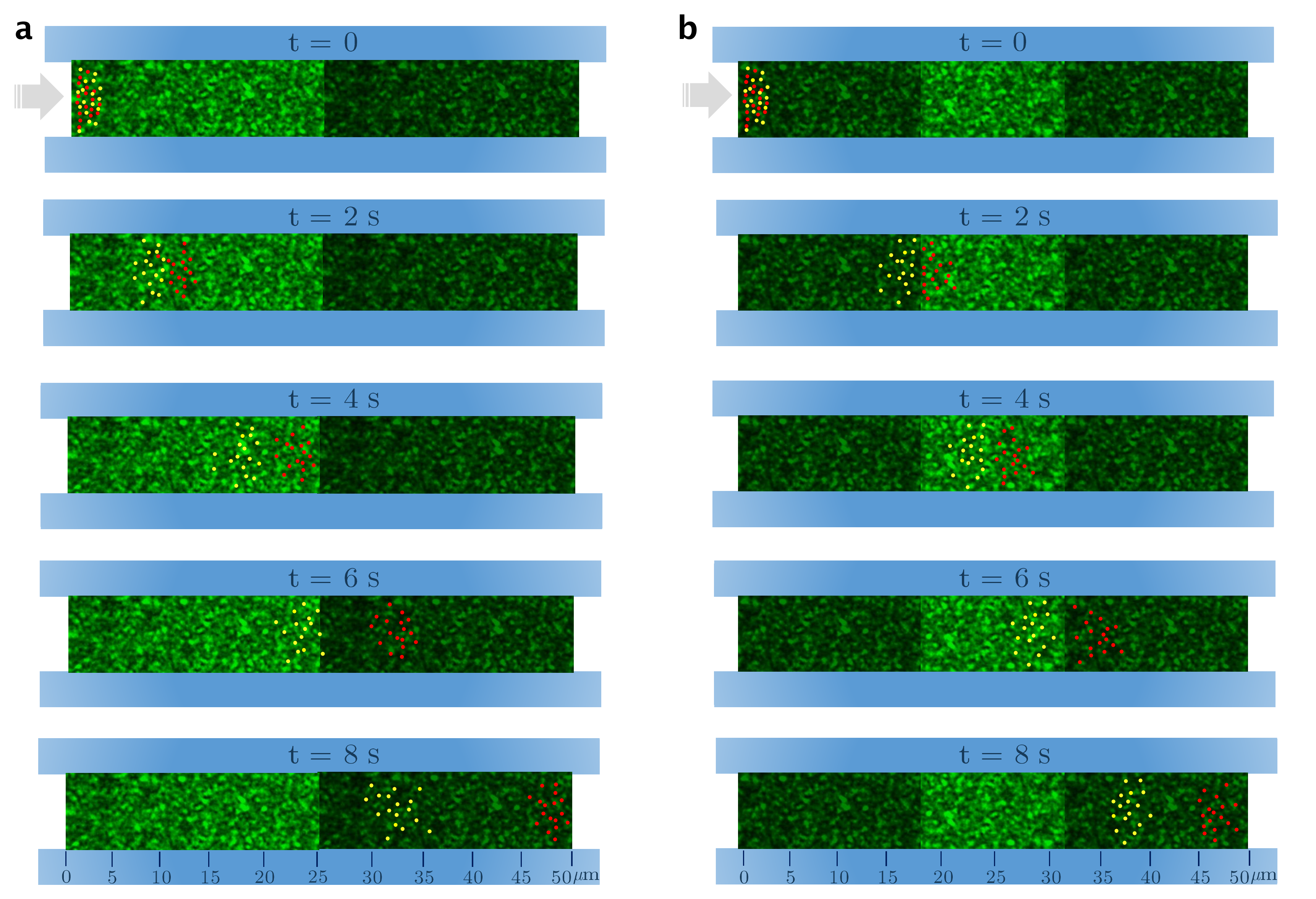}
	\caption{Microfluidic speckle filtering of NP-loaded liposome (gold color dots) from the empty ones (red dots), based on their velocity hindering in varying speckle intensity.  The average speckle intensity is changed (from $\langle I\rangle = 300$ mW/$\mu \rm{m}^2$  to $\langle I\rangle = 45$ mW/$\mu \rm{m}^2$, or vice versa) in (a) one, and (b) two boundaries. Sample time lapses belong to every 2 s and the process takes a time duration of about 8 s.}
	\label{Figure05}
\end{figure*}

We have  verified these results on simulated data considering the same  physical parameters of the experiments (Figs. 4\textbf{c} and 4\textbf{d}). The results of our simulations are in very good agreement with the results of the experiments. The motion of a Brownian particle in a static speckle field without fluid flow is the result of random thermal forces and the gradient  optical forces, which attract high-refractive-index particles towards the intensity maxima of the optical field and, on the contrary, repel the low-refractive-index ones to the intensity minima. The motion of  a Brownian particle in presence of an external potential, such as a speckle pattern,    can be modeled by solving the Langevin equation \cite{volpe2014brownian,shimoni2011multi}, in which  the inertial effects in the motion of the particle are usually neglected. Also, the influence of the particles in the speckle field is neglected because of the low  concentration of the immersed particles.  The force fields of the simulated data are obtained through the ray optics approach \cite{garcia2018sim}.  The details of the  force fields of the high- and low-refractive-index  simulated particles (Figs. 4\textbf{c} and 4\textbf{d})  moving in speckle fields show very good agreement with the experimental data.  As it is shown, the magnitude of the acting force vectors on the high-index particles is  comparable with the one of  low-index   particles  in any typical point within the speckle field, both in experimental (Figs. 4\textbf{a} and \textbf{b})and simulated  (Figs. 4\textbf{c} and \textbf{d}) force fields. 

In \cite{volpe2014brownian} G. Volpe et al.  demonstrated that by employing a time-varying speckle pattern and via continuous control of the subdiffusion (a slope smaller than 1  in the MSD($\Delta t$) function) to superdiffusion (a slope bigger than 1) transitions,   it is possible to perform optical manipulation tasks such as guiding, sieving and sorting multiple particles. Indeed, compared to periodic potentials which are characterized by  a few potential depths, ST represents a  wider possibility (due to  a distribution of intensities) to be operated for such tasks, provided that a controlling handle exists. However, controlling the variation of a produced speckle pattern comes along with challenges due to the inherent randomness of the speckle light fields. Nevertheless,  the optical forces exerted on a particle depend on the particle's physical parameters,  hence, a static speckle pattern can be employed to realize a selection sieve in the presence of flow. 
\begin{figure*}
	\includegraphics[width=\textwidth]{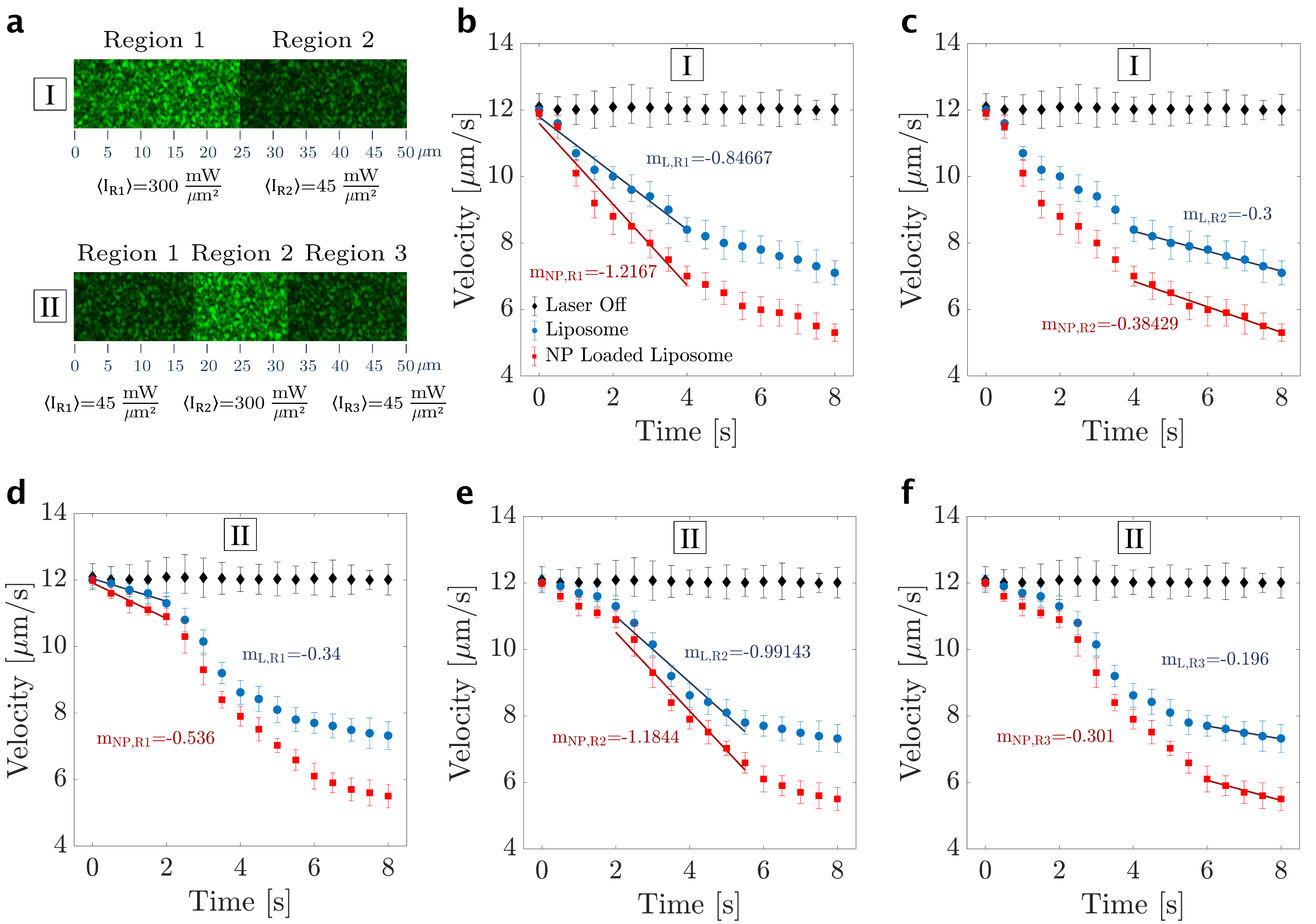}
	\caption{(a) Case \rom{1}:  the chamber surfaces is half-coated by a semi-transparent layer to reduce the transmittance of the laser light by 85\% in Region 2, and Case \rom{2}: two-third of of the chamber surface is coated (Regions 2 and 3). (b) and (c) The average velocity of empty (blue dots) and NP-loaded (red squares) liposomes and their associated fitted lines in Regions 1 (panel b) and 2 (panel c) of case \rom{1}. (d-f) The average velocity of empty  and NP-loaded liposomes and their associated fitted lines in Regions 1 (panel d), 2 (panel e) and 3 (panel f) of case \rom{2}.  The slobe of the fitted lines are in $\mu$m/s$^2$ units. The black diamonds represent the velocity of a liposome when the speckle generating laser is off, as a reference experiment.}
	\label{Figure06}
\end{figure*}

A significant  observation in this work is that  for a flow of low-index particles e.g. liposomes, the immediate effect of the presence of a speckle  field is the reduction of their drift velocities. More importantly, for the liposomes inside which we loaded  gold NPs, we observe that  the reduction of the velocities is substantially reduced. For these experiments, we fabricated special chambers in which parts of one of the building coverslips are coated by a semi-transparent thin layer, therefore, it rejects partly  the incident light. The rest of the uncoated area transmits almost all the light. In the chamber shown in Fig. 5\textbf{a} the right half of the coverslip is coated while  in the one shown in Fig. 5\textbf{b} only the central part  remains uncoated.  Then a mixture suspension of empty liposomes and NP-loaded liposomes are injected into the chamber under a constant rate via a syringe pump and the flow is live monitored. Gold color and red dots represent  the NP-loaded and empty liposomes, respectively.  In Fig. 5, for both of the two aforementioned chambers it is demonstrated that  microfluidic speckle filtering of NP-loaded liposomes from the empty ones is possible  in  a flow duration of a few seconds and within less than 50 $\mu$m range.  The separation can be done because the NP-loaded liposomes  drop behind with respect to the empty ones.
Figure 5\textbf{a} shows that the presence of  regions with different average intensities can tune the speed reduction difference, and, therefore, can boost the  separation of the two sets of particles at the boundary of the  regions. 
We demonstrate this type of control on the speed reduction difference  further in  Fig. 5\textbf{b} by adding another  boundary which enables further   control on the collective behavior of the flowing particles.  This possibility to apply  local  changes on the average intensity over the speckle field makes is possible to design useful arrangements for directed collective motion of multiple particles.   

In order to express the results in a more quantitative fashion, we track several particles  during their flow and  measure their velocities.
Figure 6 summaries the results. As shown in Fig. 6\textbf{a}, in case \rom{1} by coating a semi-transparent layer on a half of the chamber surface to reduce the transmittance of the laser light by 85\%, two regions with average speckle intensity of  $\langle I_{\rm{R1}}\rangle = 300$ mW/$\mu \rm{m}^2$  (Region 1) and  $\langle I_{\rm{R2}}\rangle = 45$ mW/$\mu \rm{m}^2$ (Region 2) are achieved. In case \rom{2} three regions of different average  speckle intensity,  $\langle I_{\rm{R1}}\rangle = 45$ mW/$\mu \rm{m}^2$ (Region 1),   $\langle I_{\rm{R2}}\rangle = 300$ mW/$\mu \rm{m}^2$  (Region 2),  and  $\langle I_{\rm{R3}}\rangle = 45$ mW/$\mu \rm{m}^2$  (Region 3) are existed. Figures 6\textbf{b,c} and 6\textbf{d-f} show the average velocity of 50 particles of each type for the first 8 s of experiments for the case \rom{1} and case \rom{2}, respectively. Blue  dots and red squares indicate the velocities of empty liposomes and NP-loaded liposomes, respectively. 
We also conduct a reference  experiment,  in which the laser is off and the speckle field is removed, therefore, the liposomes  have the constant flowing speed along with the  fluctuations  caused by  Brownian motion (black diamonds). 
It is obvious that both types of particles encounter velocity reduction. However, the reduction is substantially different for the NP-loaded and empty liposomes. The reduction is also  different in different regions. These  outcomes are  shown by the difference in the slopes of the fitted lines to the data.  In Fig. 6\textbf{b} the   lines fitted  to  velocities of the empty liposomes (blue line) and the NP-loaded liposomes (red line) in the initial seconds (moving in Region 1) have the slopes of  $m_{\rm{L,R1}}=-0.84667~\mu$m/s$^2$ and $m_{\rm{NP,R1}}=-1.2167~\mu$m/s$^2$, respectively. When  the particles cross the boundary of the two regions of case \rom{1}  (Fig. 6\textbf{c}) the decrease of the velocities of both types in Region 2  follows a different trend ($m_{\rm{L,R2}}=-0.3~\mu$m/s$^2$ and $m_{\rm{NP,R2}}=-0.38429~\mu$m/s$^2$).  For the case \rom{2} the trends of the same order in velocity reduction of empty and NP-loaded liposomes  are measured in the Regions 1 and 3, which have similar average speckle intensity ($m_{\rm{L,R1}}=-0.34~\mu$m/s$^2$, $m_{\rm{L,R3}}=-0.196~\mu$m/s$^2$ and $m_{\rm{NP,R1}}=-0.536~\mu$m/s$^2$, $m_{\rm{NP,R3}}=-0.301~\mu$m/s$^2$). However, the velocity  in Region 2  which has higher average speckle intensity decreases dramatically ($m_{\rm{L,R2}}=-0.99143~\mu$m/s$^2$ and $m_{\rm{NP,R2}}=-1.1844~\mu$m/s$^2$). 

It is obvious that more than 30\% of the velocity reduction difference between the empty and NP-loaded liposomes  in a few seconds, which is deduced from our results, is sufficient to separate out the two important classes of the particles. The separation can be followed by guiding the first outgoing particles into a specific reservoir as no NP-loaded ones. 
Proper geometry of the fluidic channels and controlled speckle field engineering can be also combined toward the application of the method in endoscopic studies. It is remarkable that the hydrodynamic interactions between dispersed particles induced with the speckle field, cause the changes in their particle-particle distances and can hinder their diffusion  \cite{segovia2019diffusion}. However,  this effect is common for either NP encapsulated or empty liposomes and can be disregarded in the presented experiment schemes.

The difference between the velocity reduction of NP-loaded and empty liposomes may be explained by the inclusion of the photophoretic forces on NPs.  Photophoretic force  is the result of thermal processes caused by the absorption of laser light  by the particles. Therefore, in the optical force calculation for metallic NPs, either entrapped inside bubbles or freely moving, the effect of the local electromagnetic fields and the photophoretic forces have to be  considered \cite{doi:10.1021/jp026731y}. The theoretical investigation of the applied forces for irregularly shaped particles or irregularly distributed laser fields can be performed  by the T-matrix method \cite{doi:10.1002/elps.200800484}.  However, quantitative consideration of the photophoretic force on an absorbing particle involves many factors, such as pressure, light properties e.g. beam profile, intensity, wavelength, and polarizations, and particle properties e.g. size, geometrical shape, and thermal conductivity \cite{shvedov2012polarization,jonavs2008light}. The effects can be  either a pushing or pulling force. It has been already shown that airborne metallic NPs may be trapped in the dark regions of speckle fields \cite{doi:10.1021/jp026731y}. Here, similarly, the photophoretic trapping  partially holds the NPs, therefore, and it hinders the associated liposomes velocities.

Gold NP, itself,  can enter liposome and  be used as a drug. Moreover, NP-loaded liposome can resemble drug-loaded containers in drug delivery, in general. 
Therefore, this platform along with easy handling of speckle fields can provide an elegant approach for controlling the drug-loaded containers, for example for collective manipulation of transfection reagents, cell markers, and carriers of molecules  \cite{lehmuskero2015laser}.

\section{Conclusions}
In conclusion, we extend the usefulness of speckle tweezers into low  refractive index micro-particles and reported different particle speckle field interaction between empty and nano-particle loaded vesicles. 
The speckle fields have great potential to collectively guide, select, trap, or push micro-particles of low- and high- refractive-index. The high-intensity regions of the speckle fields can act as a size selection sieve for dielectric particles of high-refractive-index. On the other hand, and more importantly, the low-intensity regions of the speckle fields can perform the aforementioned tasks through the size  selection sieve for dielectric particles of low-refractive-index, or through a photophoretic sieve for absorptive ones, whether embedded inside the  containers or not. 
The rather low average intensity throughout the field that illuminates  the samples can be considered as another profitable feature of the methodology. Furthermore, in addition to the inherent irregular structure of the speckle field, by proper patterning the sample chamber, it is also possible to adjust the  average intensity of the field covering an area over the sample  or apply an additional intensity  gradient, i.e., perform space-time-varying fields. Therefore, dynamic control can be applied to the samples  with the important advantages of ST  over the typical regular pattern of trapping sites. The requirement to only a few  optical components and saving physical space, insensitivity to optical aberrations, use of unfocused light, and no need for sophisticated alignment and adjustment processes, are some of the ST advantages.  These benefits make ST a simple,  cost-effective, and versatile methodology for  applications in various fields, such as  drug delivery, lab-on-a-chip, active matter, and in-vivo applications. We believe that the principal idea behind the applications of the optical ST can also be extended to acoustical trapping, which has very  recently  attracted intensive attention.

\section*{Conflicts of interest} \par 
The authors declare no conflicts of interest.

\section*{Funding} \par 
This work was partially supported by Photonics, Laser, Advanced Materials and Manufacturing Technologies Development Council of Iran, under Grant No. 97-1-33.

\section*{Acknowledgements} \par 
The authors would like to thank Saeed Mollaei and Mohammad-Reza Aghdaei for their assistance in  performing the experiments, and Bahman Farnudi for linguistic editing of the manuscript.


\end{document}